# Effect of [OH⁻] linkages on luminescent properties of ZnO nanoparticles

*Teny Theresa John[1,#], K R Priolkar[1], Aurélie Bessière[1,2], P R Sarode[1] Bruno Viana[2]*

[1]Department of Physics, Goa University, Taleigao Plateau, Goa 403206 India

[2]*Chimie Paristech, Laboratoire de Chimie Matière Condensée de Paris, UPMC, Collège de France, UMR - CNRS 7574, 11 rue Pierre et Marie Curie, 75231 Paris Cedex 05, France*
Email: krp@unigoa.ac.in

# *also at Physics Group, BITS Pilani Goa Campus, Zuarinagar 403726*

**Abstract**. Optical properties of ZnO nanoparticles prepared from a simple chemical method using sodium zincate bath show strong white light emission. X-ray absorption fine structure studies reveal a completely different local environment around Zn in these ZnO nanoparticles. The observed luminescence properties and local structural changes have been explained on the basis of a linkage between Zn and OH⁻ ions in the surface layers of ZnO nanoparticles.

## 1. Introduction

ZnO has received much attention over the past few years because it has a wide range of properties such as high transparency in the visible range, piezoelectricity, wide-band gap semiconductivity, resistivity control over the range $10^{-3}$ to $10^{5}$ ohms, large exciton binding energy (60 meV), room temperature ferromagnetism and huge magneto-optic and chemical sensing effects [1-4]. ZnO especially in its nanostructure form is currently attracting intense global interest for its photonic applications which arises from the possibility of developing low energy and environmentally friendly white light emitting technologies and laser diodes that operate above room temperature.

The luminescence spectrum of ZnO apart from exhibiting strong excitonic emission in the UV region also shows a broad band in the green and yellow region. In the literature such defects have been ascribed to various processes such as oxygen vacancy ($V_O$), interstitial zinc ($Zn_i$), singly ionized oxygen vacancy ($V^+_O$), antisite oxygen ($O_{Zn}$), zinc vacancy ($V_{Zn}$) and even oxygen interstitial ($O_i$) [5-10].

Nevertheless, the emission should be limited in order to make it a candidate for applications not only in UV but also in white light emitting systems [5, 11]. In that case the defects formation should be carefully controlled. The luminescent properties of ZnO can be further modified in its nano-size form due to high surface to volume ratio available in nanoparticles. Attempts have been made to design ZnO at the nanoscale as a means to engineer its optoelectronic properties [12-17]. Changes in physico-chemical environment of Zn in these nanoparticles or nanocomposites are considered to be the major cause of white light emission in ZnO [5-10].

In this paper we demonstrate possibility of ZnO nanoparticles, prepared by an easy solution based method, exhibiting strong visible emission without any additional dopant. We further show that these optical properties are linked to a distinctly different local structure around Zn ions in these nanoparticles with possible linkages to [OH⁻] ions.

## 2. Experimental

ZnO particles were prepared by dissolving zinc acetate in two different solutions namely ammonium hydroxide and sodium hydroxide.

The ammonium zincate bath used for the deposition of ZnO is prepared by adding ammonium hydroxide (30% ammonia solution) to an aqueous solution of zinc acetate (($CH_3COO)_2Zn.2H_2O$) of required molarity with constant stirring. At first a white precipitate of zinc hydroxide ($Zn(OH)_2$) appears. On further addition of ammonia, the precipitate dissolves forming the $(NH_4)_2ZnO_2$ bath. The final reaction that leads to the formation of ZnO from ammonium zincate bath is

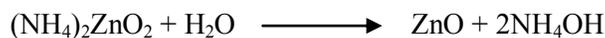

$$(NH_4)_2ZnO_2 + H_2O \longrightarrow ZnO + 2NH_4OH$$

An excess of alkali is required to have a stable ammonium zincate bath. The bath was kept at 70°C. The precipitate which formed was separated from the solution by filtration and dried in hot air oven at 50°C. It was then pressed into pellets and annealed at 160°C for 1 hr on hot plate prior to structural characterization.

The sodium zincate bath is prepared by adding sodium hydroxide (NaOH) into an equimolar solution of zinc acetate with constant stirring. The solution is then kept at room temperature for 24 hours. The final reaction leading to the formation of ZnO is

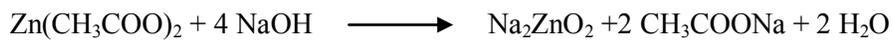

$$Zn(CH_3COO)_2 + 4\ NaOH \longrightarrow Na_2ZnO_2 + 2\ CH_3COONa + 2\ H_2O$$

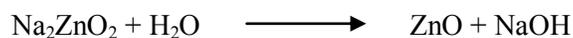

$$Na_2ZnO_2 + H_2O \longrightarrow ZnO + NaOH$$

The precipitate formed in the solution is filtered out and dried in hot air oven at 50°C. The powder is then annealed in air at different temperatures to get nanoparticles of different sizes.

The samples prepared from ammonium zincate bath and sodium zincate bath (annealed at 80°C and 300°C) are named ZnOA, ZnOn1, and ZnOn2 respectively.

XRD patterns were recorded on a Rigaku D-Max IIC diffractometer with CuK$_α$ radiation. For transmission electron microscopy (TEM) studies, an acetone dispersion of the sample was dropped onto holey carbon-coated Cu grids, and the images were recorded with FeI Technai 20 instrument at 200 kV. Fourier Transform Infrared (FTIR) spectra were recorded in vacuum (~$10^{-3}$ Torr) in the range 350 cm$^{-1}$ to 4000 cm$^{-1}$ using a FTIR-8900 (Shimadzu) spectrophotometer. For this the powdered ZnO samples were mixed with KBr powder in a ratio of 1:150 parts by weight and pressed into pellets. Optical absorption spectra were recorded on a UV-Vis 2401PC UV-Visible spectrophotometer using a integrated sphere reflectance attachment in the range 190 nm to 900 nm. The reflectance spectra were converted to absorbance using Kubelka-Munk function. Laser-excited photoluminescence spectra were recorded using the 266 nm light provided by the quadrupled frequency of a YAG:Nd laser. The emission was analyzed by a HR250 monochromator (Jobin-Yvon) coupled with an UV-enhanced intensified charge coupled device (ICCD, Roper). Extended X-ray Absorption Fine Structure (EXAFS) at the Zn K-edges were recorded at Photon Factory using beamline 7C at room temperature. For EXAFS measurements the samples to be used as absorbers, were ground to a fine powder and uniformly distributed on a scotch tape. These sample coated strips were adjusted in number such that the absorption edge jump gave $\Delta\mu t \leq 1$ where $\Delta\mu$ is the change in absorption coefficient at the absorption edge and $t$ is the thickness of the absorber. The incident and transmitted photon energies were simultaneously recorded using gas-ionization chambers as detectors. Measurements were carried out from 300 eV below the edge energy to 1000 eV above it with a 5 eV step in the pre-edge region and 2.5 eV step in the EXAFS region. At each edge, at least three scans were collected to average statistical noise. Data analysis was carried out using IFEFFIT [18] in ATHENA and ARTEMIS programs [19]. Here theoretical fitting standards were computed with FEFF6 [20,21]. The data in the k range of 3 – 14 Å$^{-1}$ and R range of 1 – 4 Å were used for analysis.

## 3. Results and Discussion

The XRD patterns obtained for the synthesized powders from different baths are shown in Figure 1a. The diffraction pattern of ZnO sample prepared using ammonium zincate bath (ZnOA) exhibits very sharp Bragg peaks indicating presence of large sized crystallites. Scanning electron microscopy studies reveal the grain size to be larger than 50 nm. The diffraction pattern of this sample is in good agreement with the pure hexagonal wurtzite structure of ZnO (space group P6$_3$mc, a = b = 3.249 Å and c = 5.206 Å, JCPDS No. 36-1451).

In the case of particles prepared using sodium zincate bath, it is seen that the positions of the diffraction peaks are in agreement with those of bulk ZnO demonstrating the formation of wurtzite nanocrystals with similar lattice parameters. Evolution of lattice parameters, cell volume as obtained from Le Bail fitting along with the grain size calculated from Scherrer formula are given in Table 1.

**Table 1 Lattice parameters, cell volume and particle size calculated using Scherrer formula from XRD data.**

| Sample | a = b (Å) | c (Å) | Cell volume (Å$^3$) | Size(nm) |
|--------|-----------|---------|---------------------|----------|
| ZnOA   | 3.251(1)  | 5.208(8)| 47.6(8)             | 47(1)    |
| ZnOn1  | 3.247(6)  | 5.213(1)| 47.6(1)             | 7(1)     |
| ZnOn2  | 3.248(2)  | 5.204(8)| 47.5(6)             | 15(1)    |

The comparison of the plots in Figure 1b makes obvious the difference in the width of the diffraction peaks arising from ZnOn1 and those from ZnOA. The broad peaks obtained for ZnOn1 arise from the finite number of diffracting planes within the finite size of the particle. In order to get more information on the size of the bulk sample and nanocrystals, the peaks in the XRD pattern were convoluted with a Lorentzian function and the full width at half maximum (FWHM) thus obtained is used to calculate grain size from the Scherrer formula given by D = 0.9 λ/β cosθ where λ is the wavelength of the x-rays and β is the FWHM of the diffracted peak at θ. Scherrer analysis reveals a reduction of crystal size from 47 nm for ZnOA to 7 nm for ZnOn1 (as can be seen from Table 1). Furthermore it may be noted from XRD pattern in Fig. 1(a) and (b) that for samples prepared from sodium zincate bath, size and perhaps the growth orientation depend on the temperature of annealing. The sample annealed at high temperatures (~300°C) (ZnOn2) shows maximum intensity peak along (101) direction as is in case of bulk ZnO, while those annealed at lower temperatures (ZnOn1) have (002) as the strongest reflection. Relative intensities of Bragg reflection in case of ZnOA exactly match with those in bulk ZnO powdered sample as given in JCPDS File No. 36-1451.

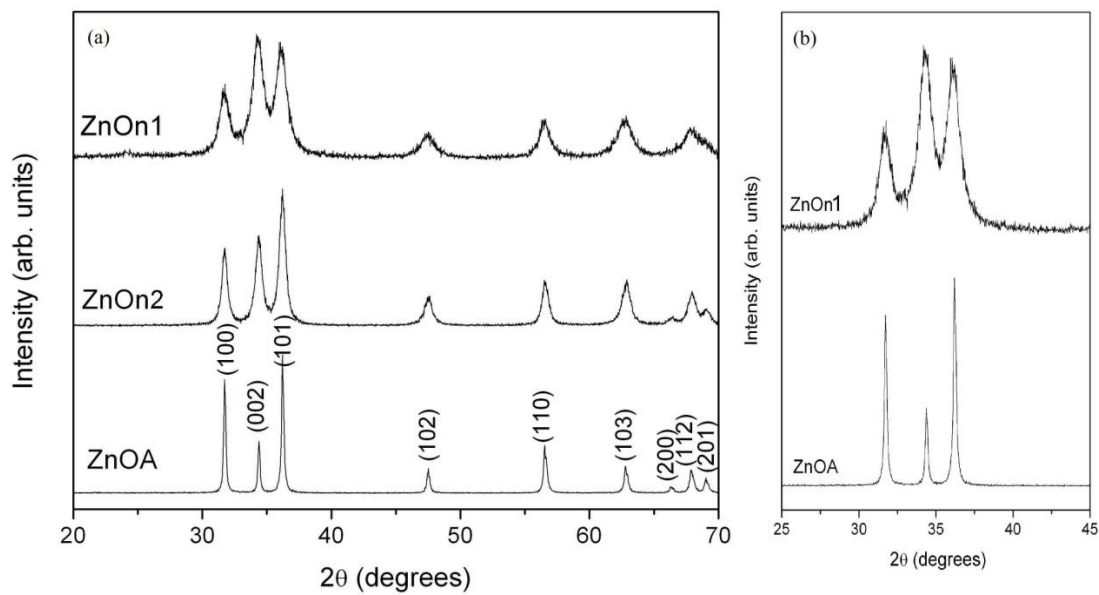

**Figure 1 (a) XRD patterns of bulk and nanoparticles from the two chemical baths, (b) Comparison of first three reflections in XRD spectra of nano particles and bulk ZnO.**

TEM images and the corresponding selected area electron diffraction (SAED) patterns for the two samples, ZnOn1 and ZnOn2 are shown in Figures 2 and 3 respectively. The images indicate that the ZnO nanoparticles consistently show crystal structure with unchanging morphology. In the case of ZnOn2 (Fig. 3), the typical morphology is spherical with an average size of about 15 nm whereas slightly elongated or elliptical particles of about 8 nm are seen in TEM image of ZnOn1 sample (Fig. 2). The changes in relative intensities of the Bragg peaks seen in XRD pattern of ZnOn1 could be due to the elongated shaped particles seen in TEM images. The nanoparticles are clearly well separated and essentially no aggregation can be found. The average particle sizes (averaged over 100 particles) obtained from TEM confirm the sizes obtained using Scherrer formula. The SAED patterns of the nanocrystals are presented in Figs. 2(b) and 3(c). The lattice spacing calculated from diffraction rings agree well with those obtained from XRD analysis. Another notable feature is that the nanoparticles especially ZnOn1 have a very narrow size distribution as can be seen from Fig. 2(c).

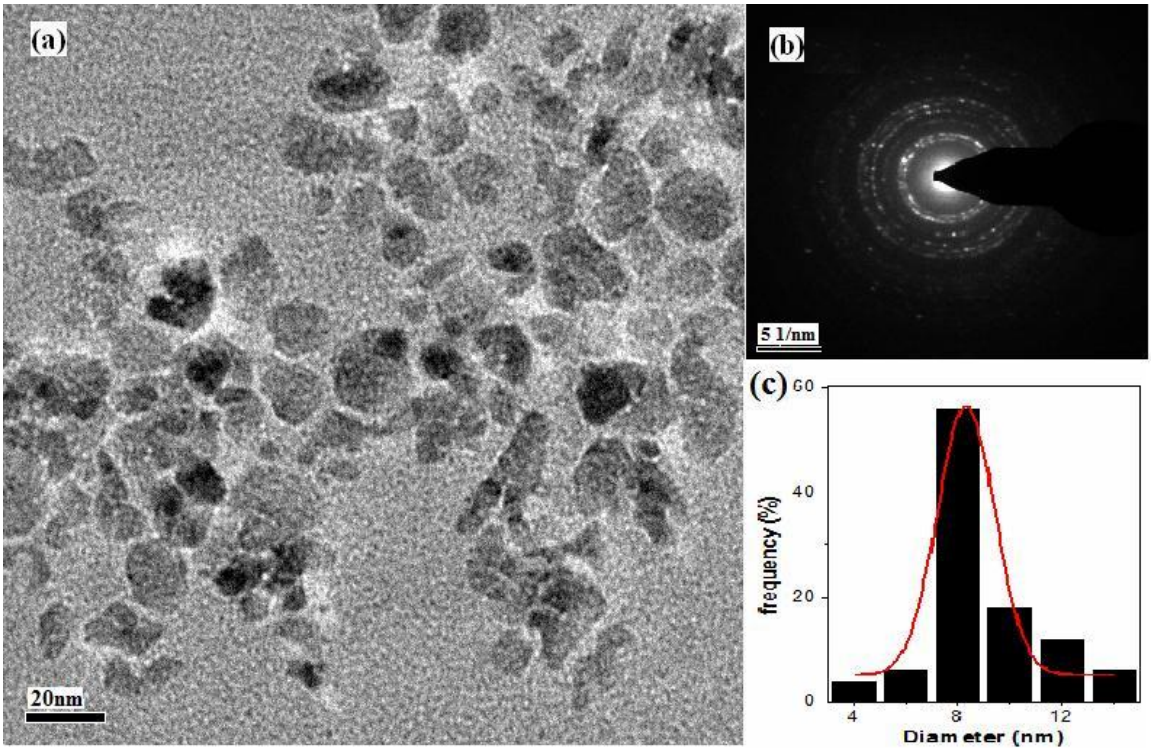

**Figure 2 (a) TEM image, (b) corresponding SAED pattern and (c) size distribution of ZnO nanoparticles (ZnOn1)**

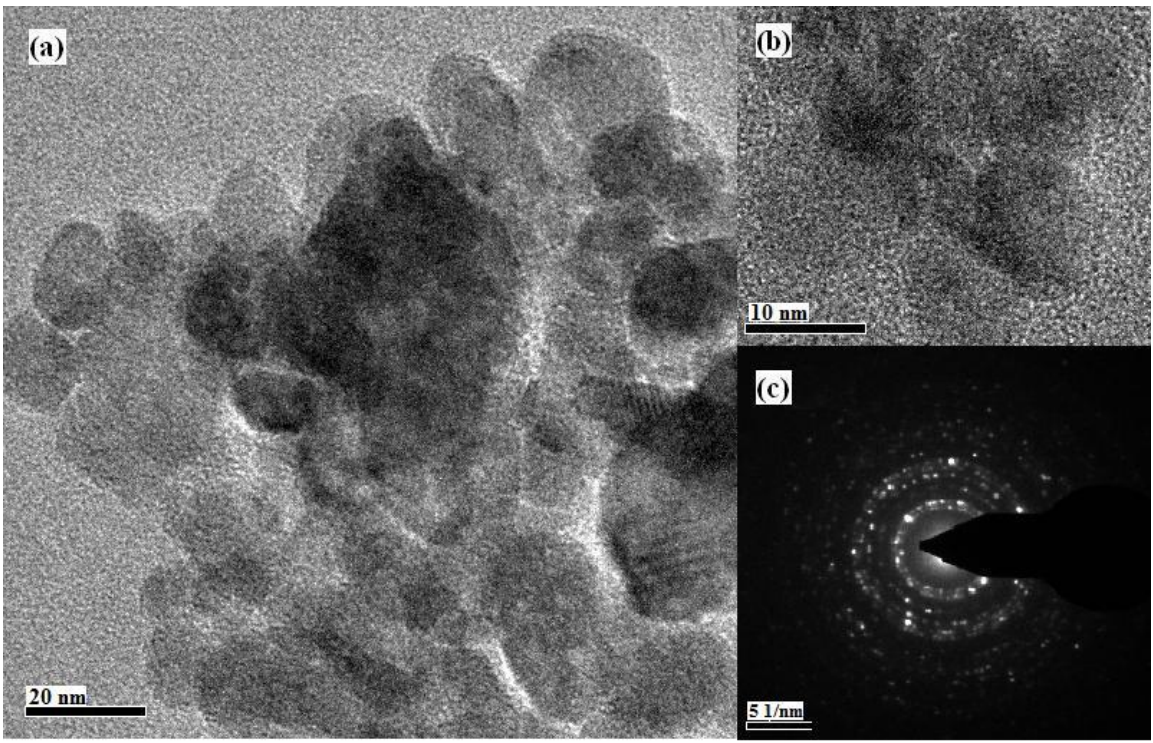

**Figure 3 (a) TEM image, (b) HRTEM image showing lattice fringes and (c) corresponding SAED pattern of ZnO nanoparticles (ZnOn2)**

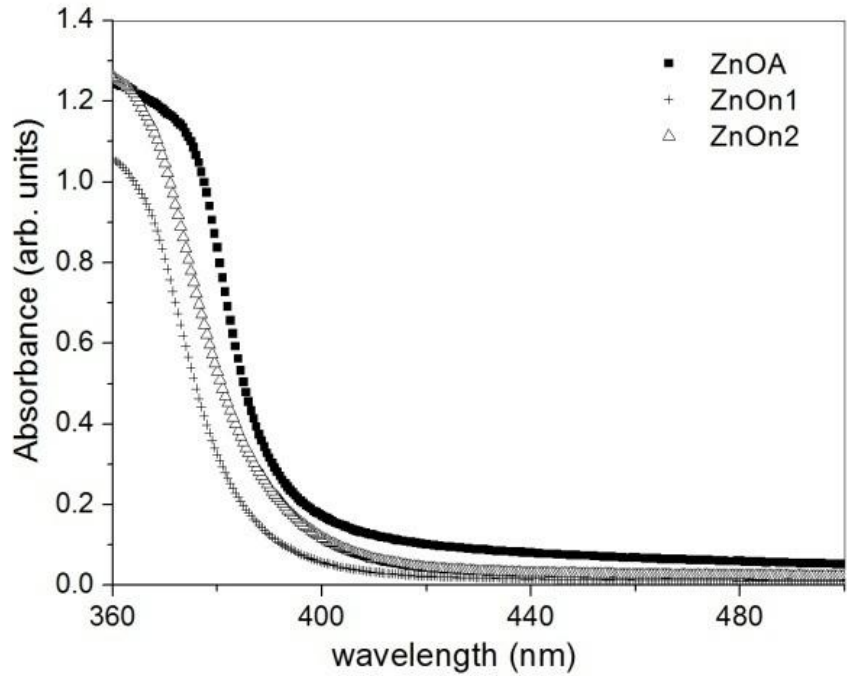

**Figure 4 UV-Vis absorption spectra demonstrating blue shift for nanoparticles**

In Figure 4 optical absorption spectra of ZnOA, ZnOn1 and ZnOn2 samples are shown. In order to obtain a precise and quantifiable measure of the shifts in the band gaps from these absorption edges, the point of inflection obtained from the minimum in the first derivative curve of the absorption spectrum is used. The band gap thus obtained for ZnOn1 corresponds to 3.33 eV (372.9 nm), indicating a blue shift of about 0.1 eV compared to the band gap 3.24 eV (381.5 nm) of ZnOA. Similar variation band gap has been observed earlier in ZnO nano particles [22, 23]. This quantum size effect can be explained qualitatively by considering a particle-in-a-box like situation where the energy separation between the levels increases as the dimensions of the box are reduced. Thus one observes an increase in the band gap of the semiconductor with a decrease in the particle size.

Figure 5 shows room temperature PL spectra recorded with an excitation of 266 nm for ZnOA, ZnOn1 and ZnOn2. All spectra show the second-order harmonic of laser line at 532 nm. The ZnOA sample presents a very sharp ultra-violet (UV) emission around 380 nm and a broad yellow emission centered at around 550 nm (~ 2.15 eV). The UV emissions are commonly attributed to the near band edge (NBE) emission that is a result of the recombination of excitons. The observation of a sharp UV emission in bulk ZnO indicates its good crystalline quality [6,23]. In addition a relatively weaker and broad emission band is seen in yellow region. The origin of this band is generally attributed to deep level defects [24-27].

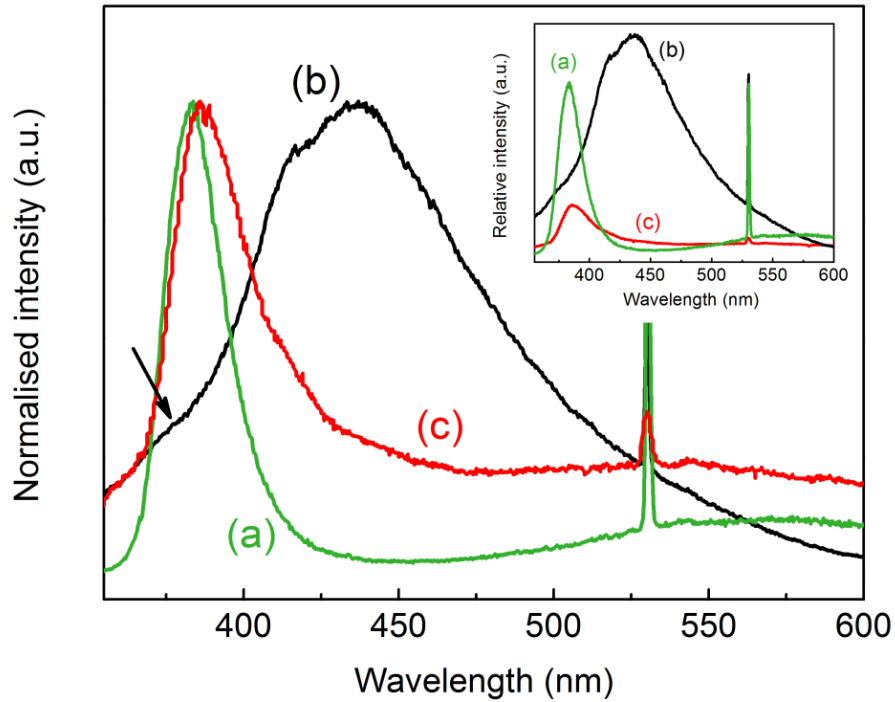

**Figure 5 Room temperature PL spectra of ZnO nanoparticles (ZnOA (a), ZnOn1 (b) and ZnOn2(c)).**

The PL spectrum of ZnOn1 sample is completely different. It shows a very strong and wide luminescence signal spanning widely over the visible spectral range. The intense luminescence signal indicates the presence of defects in the material. Some luminescence is observed in the green around 530 nm and is commonly attributed to cationic vacancies in ZnO [25, 28]. Weak excitonic luminescence at the shortest wavelength part (as indicated by an arrow in Fig. 5) of the spectrum is an indication of ultra-small particles [29]. However the most intense part of the photoluminescence is emitted as a wide band peaking at 435 nm. This is much less common in ZnO than the usual UV, green/yellow and red emissions. The origin of this blue emission in ZnO nanoparticles is not yet fully acknowledged in literature. In literature it has been attributed to carbon impurities attached to the surface of nanoparticles [30] or to interstitial oxygen ions implanted in ZnO matrix [31-33]. Furthermore such a luminescence has also been reported in nanoparticles synthesized in an excess [OH⁻] environment [22]. Although the experimental procedure does not allow for determination of quantum efficiency, it can be inferred from the relative intensities shown in the inset of Fig. 5, that luminescence in ZnOn1 is much greater than that in ZnOA. As the presence of carbon impurities would have lowered the fluorescence yield, we attribute the observed luminescence to the presence of defects in the sample. The luminescence spectrum of ZnOn2 which was subjected to higher annealing temperatures appears to be that of an intermediate between ZnOn1 and ZnOA.

In order to probe the origin for the blue emission in ZnOn1, EXAFS studies at the ZnK edge have been carried out.

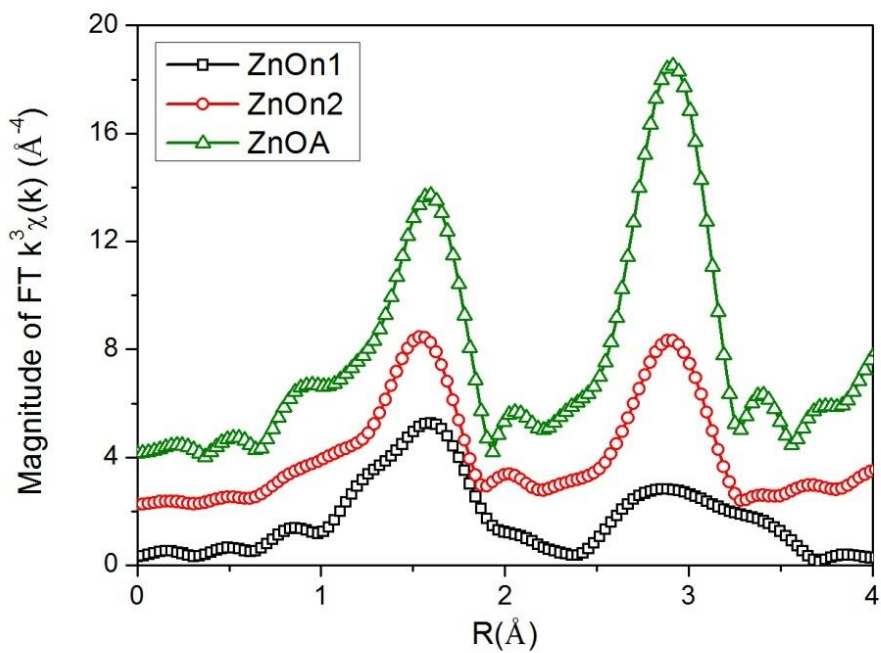

Figure 6 Magnitude of Fourier transforms of $k^3$ weighted EXAFS of Zn K-edge in ZnOA, ZnOn1 and ZnOn2

Fourier transforms (FT) of Zn K-edge EXAFS for all the three samples are presented in Figure 6. FT of ZnOA shows two distinct peaks at 1.6 Å and 2.8 Å corresponding mainly to Zn-O and Zn-Zn correlations. A fit to this spectrum with $k$-weighting = 3, in $R$-range 1.0–4.0 Å and $k$-range 3–14Å$^{-1}$ (shown as solid line in Figure 7) based on the ZnO wurtzite structure yields values of bond distances which are in agreement with those obtained from XRD studies. The parameters obtained from best fit are presented in Table II. In comparison, the FT of ZnOn2 is quite similar to that of the ZnOA sample while the FT of ZnOn1 shows striking differences. Firstly the peak corresponding to Zn-O correlation in ZnOn1 appears at slightly higher R value and is also wider as compared to that in bulk. Secondly, a clear double peak structure is observed in ZnOn1 instead of a strong well defined peak in the bulk sample at around 2.8 Å. It is therefore evident that there are local structural distortions which arise due to a decrease in particles size in ZnO. In order to extract quantitatively the changes in the local environment of Zn in ZnOn1, EXAFS data was fitted in R-space with Zn-O and Zn-Zn correlations belonging to hexagonal ZnO structure. There were several unaccounted features which were then fitted with additional Zn-O and Zn-Zn correlations. The best fit parameters are listed in Table II. It can be seen that in ZnOn1, there is an additional doubly degenerate Zn-O correlation at 2.1 Å over and above the three neighboured Zn-O bond at 2.01 Å. This Zn-O could be either due to interstitial oxygen or a hydroxide ion attached to a Zn ion on the surface. Second change in the nano sample is the breakup of 6 neighboured Zn-Zn correlations in a 4+2 configuration. This bifurcation can also be understood in terms of accommodating interstitial oxygen or a hydroxide ion in place of oxygen in the Zn neighbourhood. Thus EXAFS investigation clearly brings out the differences in local environment of Zn in ZnOn1 as compared to that in ZnOn2 and ZnOA.

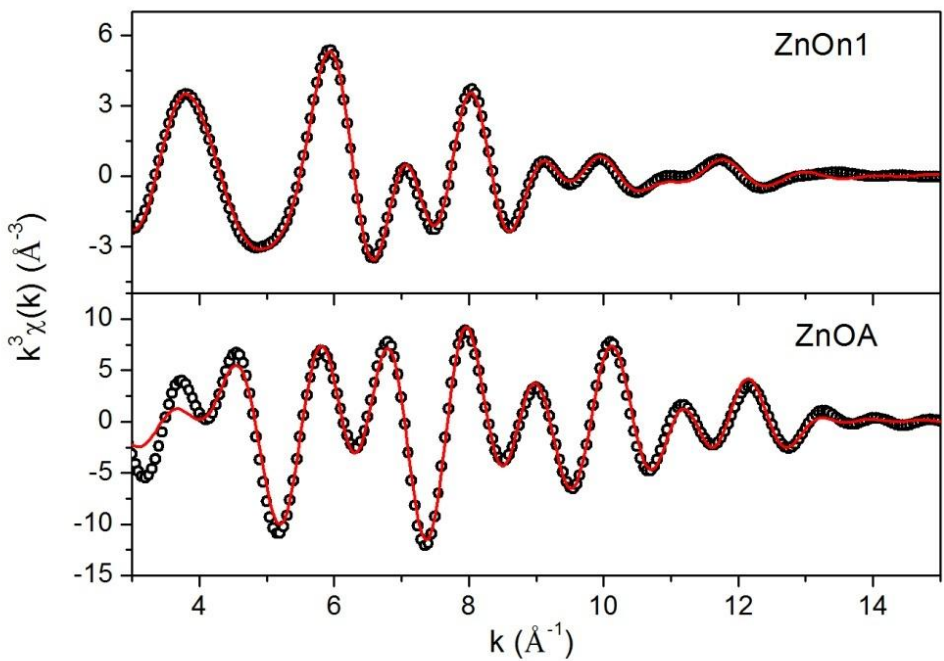

**Figure 7 Back-transformed EXAFS signal (circles) and the best fitted line in case of ZnOA and ZnOn1**

**Table II** Values of bond length, coordination number and mean square radial displacement (MSRD) in ZnOA and ZnOn1 obtained from Zn K edge EXAFS fitting. The numbers in parentheses are uncertainties in the last digit.

| Bond | Coordination Number | Bond length (Å) | MSRD (Å$^2$) |
|---|---|---|---|
| ZnOA | | | |
| Zn-O | 3 | 1.969(6) | 0.003(1) |
| Zn-Zn | 2 | 2.89(7) | 0.021(9) |
| Zn-O | 6 | 3.137(7) | 0.0012(7) |
| Zn-Zn | 6 | 3.253(4) | 0.006(1) |
| Zn-O | 6 | 3.74(2) | 0.002(2) |
| ZnOn1 | | | |
| Zn-O | 3.0(2) | 2.010(2) | 0.009(1) |
| Zn-O | 2.0(1) | 2.12(2) | 0.011(3) |
| Zn-Zn | 2.0(2) | 3.003(6) | 0.015(1) |
| Zn-O | 6.0(16) | 3.56(3) | 0.035(6) |
| Zn-Zn | 4.3(1) | 3.135(1) | 0.009(1) |
| Zn-Zn | 1.7(1) | 3.273(2) | 0.005(1) |
| Zn-O | 6.0(5) | 3.93(1) | 0.010(1) |

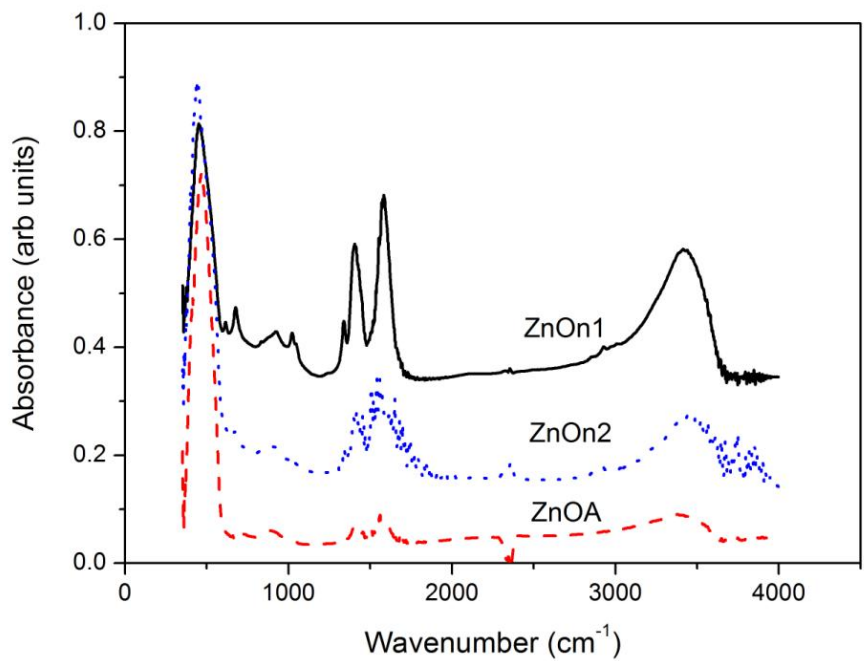

**Figure 8** FTIR spectra recorded at room temperature of (a) ZnOn1 (b) ZnOn2 (c) ZnOA in vaccum

To confirm the presence or absence of [OH$^-$] ion linkages, infra-red absorption measurements in the range 350 cm$^{-1}$ to 4000 cm$^{-1}$ have been performed. Presence of [OH$^-$] ion can be confirmed by observation of a strong stretching mode at about 3400 cm$^{-1}$ and a corresponding bending mode at about 1600 cm$^{-1}$ [35]. However, presence of these modes could also be due to presence of water adsorbed on the surface of the pellet or in the environment around the sample. Therefore the IR measurements reported here have been performed in vacuum of the order of 10$^{-3}$ mbar and on all the samples ZnOA, ZnOn1 and ZnOn2. The IR absorption spectra for these three samples are presented in Figure 8. It can be seen that very weak vibration bands corresponding to OH stretching or bending are seen in the case of ZnOA possibly due to small amount of water present in KBr powder. While quite a strong OH stretching and bending modes are present in the IR spectra of ZnOn1 and ZnOn2 samples even though all the spectra were recorded in identical conditions. This indicates that OH ions are present in these samples coming from the preparation process as these samples were prepared using Zn acetate and NaOH as starting chemicals. Some of the [OH] ions present in starting materials are linked to the Zn ion in the surface layers on ZnO particle. Such a Zn-OH linkage causes the local structure around Zn ion to change from that expected in wurtzite structure as well as be responsible for the blue emission band seen in the luminescence spectra.

## 4. Conclusion

Structural and optical properties of ZnO nanoparticles synthesized using ammonium zincate bath and sodium zincate bath have been studied. Nanoparticles prepared using sodium zincate bath (ZnOn1) shows a completely different luminescence spectrum with strong emissions in the blue and green region of visible spectrum. Based on EXAFS and infrared absorption studies it is argued that in ZnO nanoparticles synthesized at low temperatures by sodium zincate bath a linkage between Zn and [OH$^-$] ions exists in the surface states that gives rise to strong luminescence signal in the visible region. This property of multi colour emission adds another potential application for ZnO, namely white light emitting diodes (LED) by simply controlling the concentration of hydrate ions attached to Zn ions. Such a procedure does away with the use of any costly dopants such as rare-earth ions in achieving white light emission.

## Acknowledgements

Authors would like to acknowledge the financial support from Department of Science and Technology (DST), Govt. of India, New Delhi under the project No. SR/S2/CMP-42/2004 and travel support under the Utilization of International Synchrotron Radiation and Neutron Scattering facilities. Help from Ms. Asha Gupta in TEM measurements is gratefully acknowledged. Prof. M. S. Hegde is thanked for useful discussions and constant encouragements.